\documentclass[letterpaper, 10 pt, conference]{ieeeconf}  

\IEEEoverridecommandlockouts                              
\usepackage{epsfig} 
\usepackage{amsmath} 
\usepackage{amssymb}  
\usepackage{ifthen}
\usepackage{commath}
\usepackage{color}
\usepackage{cancel} 
\usepackage{cite}
\usepackage{threeparttable}
\usepackage{etoolbox}
\usepackage{mathtools}
\makeatletter
\pretocmd\start@align{%
  \if@minipage\kern-\topskip\kern-\abovedisplayskip\fi
}{}{}
\makeatother

\title{\LARGE \bf
Biomolecular resource utilization in elementary cell-free gene circuits
}

\author{Dan Siegal-Gaskins$^{1}$,  Vincent Noireaux$^{3}$, and Richard M. Murray$^{1,2}$
\thanks{For correspondence, please contact {\tt\small dsg@caltech.edu}, {\tt\small noireaux@umn.edu}, or {\tt\small murray@caltech.edu}.  Affiliations: (1) Department of Bioengineering and (2) Department of Control and Dynamical Systems, California Institute of Technology, 1200 E. California Blvd., Pasadena, CA 91125, USA; (3) School of Physics and Astronomy, University of Minnesota, 116 Church Street SE, Minneapolis, MN 55455, USA.}%
}

\begin{document}

\maketitle
\thispagestyle{empty}
\pagestyle{empty}

\begin{abstract}

We present a detailed dynamical model of the behavior of transcription-translation circuits {\it in vitro} that makes explicit the roles played by essential molecular resources.  A set of simple two-gene test circuits operating in a cell-free biochemical `breadboard' validate this model and highlight the consequences of limited resource availability.  In particular, we are able to confirm the existence of biomolecular `crosstalk' and isolate its individual sources.  The implications of crosstalk for biomolecular circuit design and function are discussed.

\end{abstract}

\section{INTRODUCTION} \label{intro}

The last several decades have witnessed significant advances in the biological sciences, in part through the application of techniques from historically distinct  areas such as mathematics, computer science, physics, and engineering.   Among the many insights that have come from this interdisciplinary approach is the existence of indirect coupling or {\it crosstalk} between biological circuit (`biocircuit') components \cite{Scott:2010cx,Mather:2010id,Chu:2011ip,Cookson:2011il,Rondelez:2012fz}.  While evidence suggests that crosstalk comes about through a shared molecular resource pool, neither this fact nor the identity of the specific resources are in general evident in the mathematical representations of biocircuits that are commonly used.  

\begin{figure}[!t]
\centering
\includegraphics[width=2in]{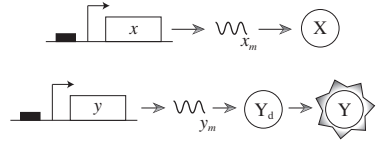}
\caption{Schematic of a simple constitutive two-gene circuit.  Noninteracting genes $x$ and $y$ are transcribed into mRNAs $x_m$ and $y_m$, which are then translated into a generic protein $\mathrm{X}$ and immature fluorescent reporter $\mathrm{Y_d}$, respectively.  $\mathrm{Y_d}$ then matures into the visible $\mathrm{Y}$.   mRNAs and proteins may also be degraded and/or diluted (not shown).}
\label{naiveCircuit}
\end{figure}

To illustrate, consider a simple biocircuit consisting of two constitutively-expressing genes, $x$ and $y$, which code for a generic protein $\mathrm{X}$ and fluorescent reporter $\mathrm{Y}$, respectively (Fig.~\ref{naiveCircuit}).  One `naive' description of the system is as follows:  $x$ is transcribed into mRNA $x_{m}$ and translated into $\mathrm{X}$, and $y$ is transcribed into mRNA $y_m$ and translated into a dark reporter protein $\mathrm{Y_d}$ that matures into the visible $\mathrm{Y}$.   All protein and mRNA species are also degraded (and/or diluted through cell division if in growing cells).  If we assume that reactions take place in sufficiently large volumes (so that stochasticity in molecule concentrations does not affect the overall dynamics), and that circuit dynamics can be approximated using mass-action kinetics, then the system may be described by the following set of ordinary differential equations (ODEs):
\begin{subequations} \label{naiveModel}
\begin{align}
\dif\, [x_{m}]/\!\dif t{}={}& k_{x,TX} [x] -  k_{xm,deg} [x_{m}] \\ 
\dif\, [\mathrm{X}]/\!\dif t{}={}& k_{x,TL} [x_{m}] - k_{X,deg}[\mathrm{X}]
\end{align}
\begin{align}
\dif\, [y_m]/\!\dif t{}={}& k_{y,TX} [y] - k_{ym,deg} [y_m] \\
\dif\, [\mathrm{Y_d}]/\!\dif t{}={}& k_{y,TL} [y_m] - (k_{mat} + k_{Y,deg}) [\mathrm{Y_d}] \\
\dif\, [\mathrm{Y}]/\!\dif t{}={}& k_{mat} [\mathrm{Y_d}] - k_{Y,deg}[\mathrm{Y}]
\end{align}
\end{subequations}
where the $k_i$ are the various reaction rates of the circuit.   While this type of model is common for the analysis of general biocircuit dynamics \cite{deJong:2002p411}, it shows no crosstalk or molecular competition effects---the model circuit output $[\mathrm{Y}]$ is completely unaffected by the presence of $x$.

This particular model formalism assumes that the essential transcription and translation (TX-TL) machinery, including transcription initiation factors, RNA polymerase (RNAP), and ribosomes, exist in sufficiently high concentrations and that their utilization by one component has no noticeable effect on others.   Clearly, for the study of crosstalk, an approach that does not rely on these assumptions is required.    To this end, various theoretical frameworks have recently been developed (e.g., \cite{Yeung:2012tz,DeVos:2011ba}); however, models that (i) may be used to explore resource utilization effects, (ii) distinguish between different sources, and (iii) are supported by experimental data are still lacking.

We developed a detailed mathematical model of {\it in vitro} TX-TL circuits that consist of only two genes, with a level of complexity sufficient to capture effects that may arise via the sharing of fixed-concentration molecular resources.   An important design criterion was that the model have a general form that could be easily expanded to more complex circuits, and that the individual sources of crosstalk and their relative contributions to the total could be identified with a small number of simple experiments.  The resultant model is shown in Section \ref{model}, with the experimental `breadboard' used to validate it described in Section \ref{breadboard}.  The simulated and experimental results identifying the sources of biomolecular crosstalk are presented in Section \ref{results}, along with predictions as to how the level of crosstalk may be affected by resource and DNA concentrations and binding affinities.  In Section \ref{sigma} we show additional results for the special case when the second gene is a constitutively-expressing alternative sigma factor.  A discussion of all these results is given in Section \ref{discussion}.

\section{DETAILED MODEL FOR CONSTITUTIVE EXPRESSION OF TWO GENES IN VITRO} \label{model}

For the two-gene system described above, a more detailed model for {\it in vitro} operation that makes explicit the role of TX-TL machinery is

\vspace{.2cm}
\begin{minipage}[t]{0.45\linewidth}
    \begin{align*}
\mathrm{E} + \mathrm{S1} & \rightleftharpoons \mathrm{ES1} \\
\mathrm{ES1} + x & \rightleftharpoons x\!\!:\!\!\mathrm{ES1} \\
\mathrm{ES1} + y & \rightleftharpoons y\!\!:\!\!\mathrm{ES1} \\
x\!\!:\!\!\mathrm{ES1} \rightarrow x & + \mathrm{ES1} + x_{m} \\
y\!\!:\!\!\mathrm{ES1} \rightarrow y & + \mathrm{ES1} + y_{m} \\
x_{m} & \rightarrow \varnothing \\
y_{m} & \rightarrow \varnothing
    \end{align*}
\end{minipage}
\begin{minipage}[t]{0.45\linewidth}
    \begin{align*}
\mathrm{R} + x_{m} & \rightleftharpoons x_{m}\!\!:\!\!\mathrm{R} \\
\mathrm{R} + y_{m} & \rightleftharpoons y_{m}\!\!:\!\!\mathrm{R} \\
x_{m}\!\!:\!\!\mathrm{R} \rightarrow x&_{m} + \mathrm{R} + \mathrm{X} \\
y_{m}\!\!:\!\!\mathrm{R}  \rightarrow y&_{m} + \mathrm{R} + \mathrm{Y_d} \\
\mathrm{Y_d} & \rightarrow \mathrm{Y} \\
\mathrm{X} & \rightarrow \varnothing \\
\mathrm{Y_d} & \rightarrow \varnothing \\ 
\mathrm{Y} & \rightarrow \varnothing \  .
    \end{align*}
\end{minipage}
\vspace{.2cm}

\noindent where $\mathrm{E}$, $\mathrm{R}$, $\mathrm{S1}$, and $\mathrm{ES1}$ represent free core RNAP, free ribosome, the primary sigma factor (necessary for transcription initiation), and the primary sigma factor-RNAP holoenzyme, respectively.  RNAP holoenzymes bound to DNA and ribosomes bound to mRNA transcripts are represented by ($:$).  The ODEs for this expanded model are:


\vspace{-.3cm}
\begin{subequations} \label{modeleqs}
{\footnotesize
\begin{align}
\dif\, [x_{m}]/\!\dif t{}={}& k_{x,TX} [x\!\!:\!\!\mathrm{ES1}] - k_{xm,deg} [x_{m}] \nonumber \\
&-k_{\mathrm{X}+} [\mathrm{R}] [x_{m}] + (k_{\mathrm{X}-}+k_{x,TL})[x_{m}\!\!:\!\!\mathrm{R}] \\
\dif\, [y_m]/\!\dif t{}={}& k_{y,TX} [y\!\!:\!\!\mathrm{ES1}] - k_{ym,deg} [y_m] \nonumber \\
&-k_{\mathrm{Y}+} [\mathrm{R}] [y_{m}] + (k_{\mathrm{Y}-}+k_{y,TL})[y_{m}\!\!:\!\!\mathrm{R}] \\
\dif\, [x_{m}\!\!:\!\!\mathrm{R}]/\!\dif t{}={}& k_{\mathrm{X}+} [\mathrm{R}] [x_{m}] - (k_{\mathrm{X}-}+k_{x,TL})[x_{m}\!\!:\!\!\mathrm{R}] \\
\dif\, [y_{m}\!\!:\!\!\mathrm{R}]/\!\dif t{}={}& k_{\mathrm{Y}+} [\mathrm{R}] [y_{m}] - (k_{\mathrm{Y}-}+k_{y,TL})[y_{m}\!\!:\!\!\mathrm{R}] \\
\dif\, [\mathrm{X}]/\!\dif t{}={}& k_{x,TL} [x_{m}\!\!:\!\!\mathrm{R}] - k_{X,deg} [\mathrm{X}] \label{2ndProteinRate} \\
\dif\, [\mathrm{Y_d}]/\!\dif t{}={}& k_{y,TL} [y_{m}\!\!:\!\!\mathrm{R}] - (k_{mat} + k_{Y,deg}) [\mathrm{Y_d}] \\
\dif\, [\mathrm{Y}]/\!\dif t{}={}& k_{mat} [\mathrm{Y_d}] - k_{Y,deg}[\mathrm{Y}] \\
\dif\, [\mathrm{ES1}]/\!\dif t{}={}& k_{\mathrm{ES1}+}[\mathrm{E}]\big([\mathrm{S1}]_\text{tot}-[\mathrm{ES1}]\big)-k_{\mathrm{ES1}-}[\mathrm{ES1}] \nonumber \\
&-k_{x+} [\mathrm{ES1}]\big([x]_\text{tot}-[x\!\!:\!\!\mathrm{ES1}]\big) \nonumber \\
&+(k_{x-}+k_{x,TX})[x\!\!:\!\!\mathrm{ES1}]  \nonumber \\
&-k_{y+} [\mathrm{ES1}]\big([y]_\text{tot}-[y\!\!:\!\!\mathrm{ES1}]\big) \nonumber \\
&+(k_{y-}+k_{y,TX})[y\!\!:\!\!\mathrm{ES1}] \\
\dif\, [x\!\!:\!\!\mathrm{ES1}]/\!\dif t{}={}& k_{x+} [\mathrm{ES1}]\big([x]_\text{tot}-[x\!\!:\!\!\mathrm{ES1}]\big)-k_{x-}[x\!\!:\!\!\mathrm{ES1}]  \nonumber \\
&+k_{x,TX}[x\!\!:\!\!\mathrm{ES1}] \\
\dif\, [y\!\!:\!\!\mathrm{ES1}]/\!\dif t{}={}& k_{y+} [\mathrm{ES1}]\big([y]_\text{tot}-[y\!\!:\!\!\mathrm{ES1}]\big)-k_{y-}[y\!\!:\!\!\mathrm{ES1}] \nonumber \\
&+k_{y,TX}[y\!\!:\!\!\mathrm{ES1}]
\end{align}
}%
\end{subequations}
with conservation relations
\begin{subequations}  
{\small 
\begin{gather}
[\mathrm{E}] {}={}[\mathrm{E}]_\text{tot}-[x\!\!:\!\!\mathrm{ES1}]f(x)-[y\!\!:\!\!\mathrm{ES1}]f(y) - [\mathrm{ES1}] \label{totalE} \\
\shortintertext{\normalsize and} 
[\mathrm{R}] {}={}[\mathrm{R}]_\text{tot}-[x_{m}\!\!:\!\!\mathrm{R}]g(x)-[y_{m}\!\!:\!\!\mathrm{R}]g(y) \  .
\end{gather} 
}%
\end{subequations} 
$[\mathrm{E}]_\text{tot}$, $[\mathrm{R}]_\text{tot}$, $[\mathrm{S1}]_\text{tot}$, $[y]_\text{tot}$, and $[x]_\text{tot}$ represent the fixed total concentrations of species in the reaction volume, and factors of the form $f(a)=1+k_{a,TX}(L_a/V_{TX})$ and $g(a)=1+k_{a,TL}(L_a/V_{TL})$ account for the loading of multiple holoenzymes and ribosomes on the gene and mRNA templates \cite{mathews2007translational}.  $L_a$ is the length (in bp) of gene $a$ and $V_{TX}$ and $V_{TL}$ represent the rates of progression (in nucleotides per second) of RNAP along the DNA and ribosome along the mRNA, respectively.   

It is worth noting that, in contrast with the more common type of biocircuit model that assumes the validity of the Michaelis-Menten kinetics approximation \cite{Segel:1988vr}, we have made no assumptions as to the timescales of various reactions or the relative concentrations of reacting species.    

\section{A CELL-FREE `BREADBOARD' FOR BIOCIRCUIT TESTING} \label{breadboard}

Experimental verification of biocircuit models such as this is often challenging, due in part to the complexity of the systems and the context-dependence of their components (see, e.g., \cite{Marguet:2010ko,Tan:2009de,Cardinale:2012jh}).  As a result there has been considerable interest in using simple {\it in vitro} platforms for circuit development,  characterization, and model validation \cite{Hockenberry:2012jp}.  Important steps have been made in recent years with the development of a TX-TL `breadboard': an {\it in vitro} system that allows TX-TL processes to take place using molecular machinery extracted from {\it E. coli} \cite{Shin:2010p1243, Shin:2012vj}.  Endogenous DNA and mRNA from the cells is eliminated  so that biocircuits of interest may be studied in isolation with no other genetic material present.   The breadboard also allows for tight control over reaction conditions and the concentrations of circuit components---control which is difficult to achieve {\it in vivo}.  It is thus an ideal environment for establishing the validity of biocircuit models in general and for confirming the existence of crosstalk in simple genetic circuits. 

\section{RESULTS} \label{results}

In our cell-free system, protein species are stable against degradation and there is no dilution through cell division.  We thus set $k_{X,deg}=k_{Y,deg}=0$.   Under these conditions, the model \eqref{modeleqs} predicts the existence of a time $T$ after which the fluorescent protein concentration increases linearly; i.e., $[\mathrm{Y}] \propto t$ for $t>T$.

\begin{figure}[thpb]
\centering
\includegraphics[width=2.25in]{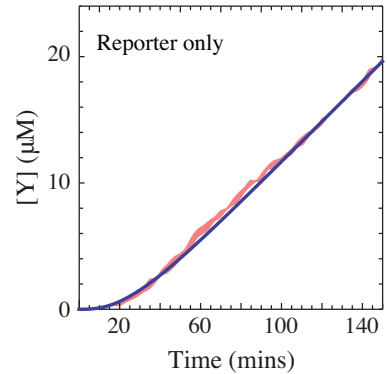}
\caption{Expression of a fluorescent reporter $y$ in the TX-TL breadboard system, with $[y]_\text{tot}=2$ nM.  Solid line is the result of simulation, and shaded area indicates the standard deviation (n=2) of reporter concentration as determined by fluorescence.}
\label{reporterData}
\end{figure}

\begin{figure}[thpb]
\centering
\includegraphics[width=2.75in]{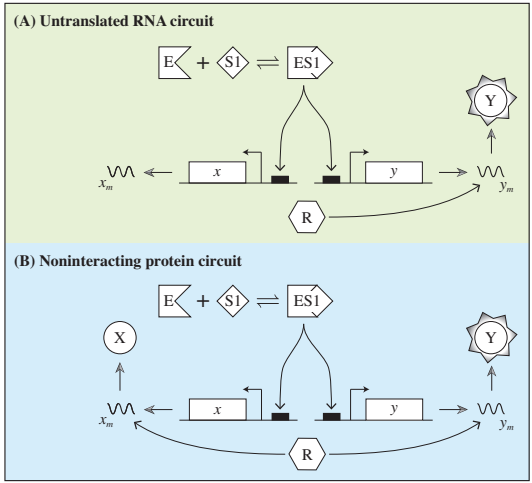}
\caption{Schematics of tested circuits. Genes $x$ and $y$, driven by constitutive promoters (filled rectangles), are transcribed into RNAs $x_m$ and $y_m$.   $x_m$ is either (A) untranslated or (B) translated into a generic noninteracting protein $\mathrm{X}$.  $y_m$ is translated into an immature fluorescent reporter $\mathrm{Y_d}$ which matures into the visible $\mathrm{Y}$.  Additional arrows represent complex formation and the regulatory roles of various molecular species as described in the text.}
\label{circuits}
\end{figure}

When only the fluorescent reporter gene is present ($[x]_\text{tot}=0$ and $[y]_\text{tot}=2$ nM), both simulation and experiment show the expected linearly-increasing fluorescent protein concentration (for $t>T$) and are in good agreement (Fig.~\ref{reporterData}).  The behavior of the system when two genes are present is less easily predictable.  It is reasonable to suspect that RNAP ($\mathrm{E}$) and ribosomes ($\mathrm{R}$) contribute to total crosstalk, since an increase in the concentration of $x$ could result in a sequestration of $\mathrm{ES1}$ away from $y$, and an increase in the amount of $x_{m}$ could decrease the amount of free $\mathrm{R}$ available to translate $y_{m}$.  We investigated---computationally and experimentally---two different circuits designed to test for RNAP and ribosome utilization effects and to distinguish between them:  in one case, $x$ encodes a small untranslated RNA to which there is no ribosomal binding (Fig.~\ref{circuits}A), and in the other, it encodes a protein that has no direct interactions with $y$ (Fig.~\ref{circuits}B).   

\subsection{Untranslated RNA circuit: simulated and experimental results}

The use of an untranslated RNA molecule allows us to determine the contribution of RNAP alone to crosstalk (Fig.~\ref{circuits}A).  In simulations, the rate of association of $\mathrm{R}$ to $x_m$, $k_{\mathrm{X}+}$, is set to zero.   At two low but biologically-relevant concentrations, $[x]_\text{tot}=0.1$ nM and $[x]_\text{tot}=1$ nM, we find no discernible effect on the rate of production of $\mathrm{Y}$; both simulated functions are linear (for $t>T$) and overlaid and consistent with experiment (Fig.~\ref{dataAB}A).   Clearly, when the additional DNA is present at low concentrations, and for this particular set of rate constants and binding affinities, the crosstalk introduced by RNAP holoenzyme alone is negligible.

\subsection{Noninteracting protein circuit: simulated and experimental results}

\begin{figure}[thpb]
\centering
\includegraphics[width=3.5in]{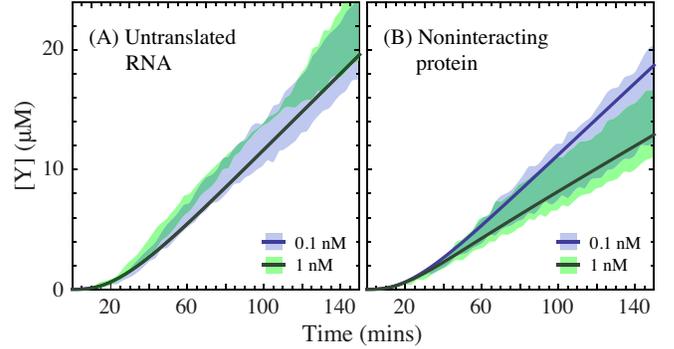}
\caption{Simulation and experimental results for implementations of the circuits schematized in Fig.~\ref{circuits}, in which the second gene encodes (A) a small untranslated RNA, and (B) a noninteracting protein.  Solid lines are simulation results, and shaded areas indicate the standard deviation (n=2) of reporter concentration as determined by   fluorescence for $[x]_\text{tot}=0.1$ nM and $[x]_\text{tot}=1$ nM.}
\label{dataAB}
\end{figure}

We now consider the effect of ribosome sequestration on circuit output when the second gene codes for a noninteracting protein  (Fig.~\ref{circuits}B).  We use the full model of Section~\ref{model} with all rates and concentrations positive.  As with the single-gene control and untranslated RNA circuit, simulations show a linear increase in output for $t>T$;  however, the model predicts a slope for $\dif\, [\mathrm{Y}]/\!\dif t$ that is different for $[x]_\text{tot}=0.1$ nM and $[x]_\text{tot}=1$ nM (Fig.~\ref{dataAB}B), and the experimental data is consistent with this prediction.  Thus, unlike RNAP, ribosomes appear to be a limiting resource and that even low levels of auxiliary ribosome targets can lead to a reduction in the circuit output.

\subsection{Sensitivity of output to changes in RNAP and ribosome concentrations} 

\begin{figure}[bt]
\centering
\includegraphics[width=3.43in]{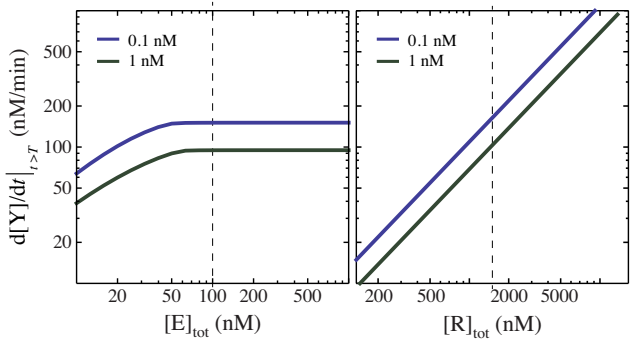}
\caption{Fluorescent protein production rate for $t>T$ as a function of total core RNAP (left) and ribosome (right) concentrations, for $[x]_\text{tot}=0.1$ nM and $[x]_\text{tot}=1$ nM.  Dashed lines indicate nominal values of $[\mathrm{E}]_\text{tot}$ and $[\mathrm{R}]_\text{tot}$ in breadboard environment.}
\label{rateVsResources}
\end{figure}

Limited resources can have a significant effect on the robustness of even simple circuits \cite{Stelling:2004ed}, a fact that is true both {\it in vivo} and {\it in vitro}.  However, given the hard limits on resource concentrations in cell-free environments (as compared with {\it in vivo} systems, in which the levels of RNAP molecules and ribosomes are regulated to some degree by the cell \cite{Nomura:1999wa,Shepherd:2001jj}), the potential for adverse limit-related effects is amplified.   We thus used our full noninteracting protein model to determine the sensitivity of the output to changes in the concentration of total core RNAPs $[\mathrm{E}]_\text{tot}$ and ribosomes $[\mathrm{R}]_\text{tot}$.  We find that $\dif\, [\mathrm{Y}]/\!\dif t$ is completely insensitive to changes in RNAP concentration when RNAP levels are high (Fig.~\ref{rateVsResources}, left).  Interestingly, the system naturally operates in this regime with the nominal value of $[\mathrm{E}]_\text{tot}$.  On the other hand, the total concentration of ribosomes has a significant effect on $\dif\, [\mathrm{Y}]/\!\dif t$,  one that increases dramatically with increasing $[\mathrm{R}]_\text{tot}$ (Fig.~\ref{rateVsResources}, right).

\subsection{Gene concentrations, binding affinities, and resource limits} 

\begin{figure}[bt]
\centering
\includegraphics[width=3.5in]{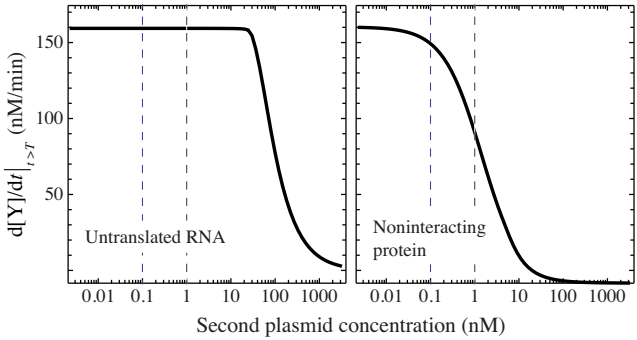}
\caption{Fluorescent protein production rate for $t>T$ as a function of the concentrations of the second gene when it codes for an untranslated RNA (left panel) or a typical noninteracting protein (right panel).  Dashed lines indicate the concentrations $[x]_\text{tot}$ used elsewhere in this work.}
\label{rateVsConc}
\end{figure}

We may also use our model to answer questions about the system that would be difficult to address experimentally.  These include determining the level of additional genes above which RNAP becomes a limiting resource, and below which the ribosomal loading does not lead to any significant crosstalk.  In Fig.~\ref{rateVsConc} we see how $\dif\, [\mathrm{Y}]/\!\dif t$ is affected by the concentration of a second gene over 6 orders of magnitude.  As before, we use the simulated untranslated-RNA gene (with $k_{\mathrm{X}+}=0$) to isolate and predict the effect of holoenzyme utilization.  We find that it is only at an additional gene concentration of $\sim$25 nM (Fig.~\ref{rateVsConc}, left), or 12.5X the 2 nM total reporter concentration, that crosstalk arising from limited holoenzyme appears as a $>$1\% change in the output.  On the other hand, ribosome-related crosstalk begins to manifest itself (as a $>$1\% change in the output) at concentrations as low as 0.75\% of $[y]_\text{tot}$, or $\sim$15 pM (Fig.~\ref{rateVsConc}, right).    

\begin{figure}[bt]
\centering
\includegraphics[width=2.25in]{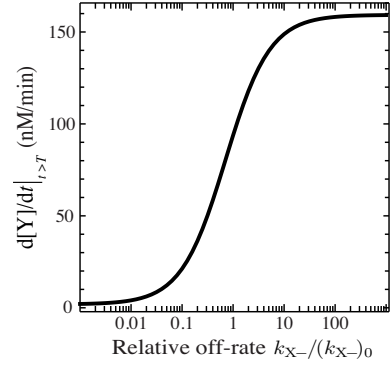}
\caption{Fluorescent protein production rate for $t>T$ as a function of the strength of the second gene's ribosomal binding site relative to that of the fluorescent reporter.  $[x]_\text{tot}=1$ nM.}
\label{rateVsOff}
\end{figure}

While crosstalk may be commonplace in natural circuits, we are not limited to naturally-occurring parts when constructing new biocircuits;  synthetic biological tools allow us to adjust many properties of circuit components, including degradation rates and resource binding affinities.  Our model may thus be used as a circuit design aid, to predict, for example, how much the ribosomal binding off-rate must be modified in order to ameliorate the effects of ribosomal crosstalk.  With $[x]_\text{tot}=1$ nM and all other parameters held fixed, our model predicts that a greater than 50-fold increase or decrease in $k_{\mathrm{X}-}$ is needed to eliminate the ribosome loading effects (Fig.~\ref{rateVsOff}).  As might be expected, a significant increase in $k_{\mathrm{X}-}$ produces an output that is little different from that of the single reporter control, while a significant decrease reduces the output to near zero as all ribosomes are sequestered by $x_m$.  Interestingly, we find that the sensitivity of the output to the off-rate parameter (as determined by the slope of $\dif\, [\mathrm{Y}]/\!\dif t$ vs. $k_{\mathrm{X}-}$) is highest at the natural value of $k_{\mathrm{X}-}$, but that this sensitivity is relatively unchanged over two orders of magnitude.  The effects of variation in other circuit parameters may be similarly tested. 

\section{SPECIAL CASE: ALTERNATIVE SIGMA FACTORS} \label{sigma}

\begin{figure}[thpb]
\centering
\includegraphics[width=2.75in]{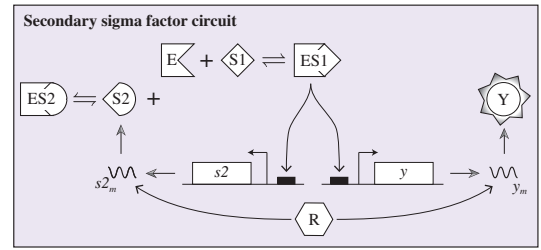}
\caption{Schematic showing molecular interactions when a secondary sigma factor is present.  Symbols are as in Fig.~\ref{circuits}.}
\label{sigmaFactorCircuit}
\end{figure}

Certain classes of proteins may contribute crosstalk effects in addition to those introduced by ribosome sequestration;  for example, alternative sigma factors that compete for access to free core RNAP and thus lead to a reduction in activity from orthogonal sigma factor-specific promoters (Fig.~\ref{sigmaFactorCircuit}).    Experimental evidence supporting sigma factor sequestration has been found {\it in vivo} \cite{Farewell:1998wc} and using purified sigma factor subunits \cite{Maeda:2000va}.   The question of the effect of alternative sigma factors on circuit performance is an important one given their relevance to complex biocircuit design:  the promoter selectivity that sigma factors confer to RNAP \cite{Osterberg:2011bk} can lead to an increase in the number of available transcriptional control elements, beyond the standard library of repressors and activators now commonly used. 

\begin{figure}[thpb]
\centering
\includegraphics[width=2.25in]{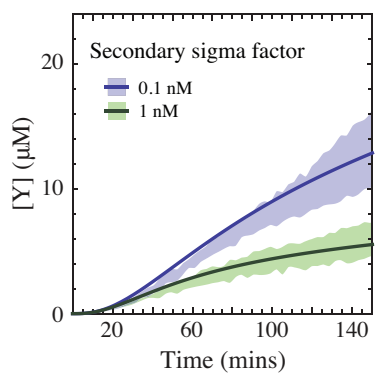}
\caption{Simulation and experimental results for implementation of the secondary sigma factor circuit schematized in Fig.~\ref{sigmaFactorCircuit}.  Solid lines are simulation results, and shaded areas indicate the standard deviation (n=2) of reporter concentration as determined by fluorescence for $[s2]_\text{tot}=0.1$ nM and $[s2]_\text{tot}=1$ nM.}
\label{dataC}
\end{figure}

In order to determine if our model formalism predicts additional resource-loading-type effects when a secondary constitutively-expressed sigma factor is introduced to the system, we add the following equation to the model \eqref{modeleqs}:
{\small
\begin{align}
\dif\, [\mathrm{ES2}]/\!\dif t{}={}& k_{\mathrm{ES2}+}[\mathrm{E}][\mathrm{S2}] -k_{\mathrm{ES2}-}[\mathrm{ES2}] \ ,
\end{align}
}%
and modify Eqs.~\eqref{2ndProteinRate} and \eqref{totalE} to be
{\small
\begin{align}
\dif\, [\mathrm{S2}]/\!\dif t{}={}&k_{s2,TL} [s2_{m}\!\!:\!\!\mathrm{R}] - k_{\mathrm{ES2}+}[\mathrm{E}][\mathrm{S2}] + k_{\mathrm{ES2}-}[\mathrm{ES2}]
\end{align}
}%
and
{\footnotesize
\begin{align}
[\mathrm{E}]{}={}&[\mathrm{E}]_\text{tot}-[s2\!\!:\!\!\mathrm{ES1}]f(s2)- [y\!\!:\!\!\mathrm{ES1}]f(y) - [\mathrm{ES1}] - [\mathrm{ES2}]\ ,
\end{align}
}%
respectively. (Notationally, references to `$x$' and `$\mathrm{X}$' have been replaced with `$s2$' and `$\mathrm{S2}$'.)  Simulated and experimental results are shown in Fig.~\ref{dataC}.    We note that (1) core RNAP sequestration by a secondary sigma factor has a more pronounced effect on the circuit output than does the ribosome loading, and particularly at higher gene concentrations, and (2) this sequestration, unlike the untranslated RNA and noninteracting protein cases, results in fluorescent protein production rates that are sublinear for the 2.5 hours of the experiment.

\section{DISCUSSION} \label{discussion}

We have presented a detailed model for a simple two-gene regulatory biocircuit operating {\it in vitro} that makes explicit the important functional roles played by RNA polymerase, sigma factors, and ribosomes and that provides insight into how these resources are shared between components.   This model, with support from our cell-free experiments, demonstrates that even a single noninteracting protein-coding gene added at a low concentration can introduce significant crosstalk through ribosomal loading.  Additional simulations suggest that the performance of the circuit is insensitive to changes in RNAP concentration but highly sensitive to ribosome concentration at physiologically-relevant levels of component DNA.  We also show that ribosome utilization effects may be difficult to avoid in any natural circuit of even minimal complexity; an elimination of these effects would require either an exceedingly low level of circuit DNA or a substantial modification of the ribosome binding affinities.  Lastly, we show through a simple extension of the model and supporting experiments that the presence of a constitutively-expressing alternative sigma factor gene decreases circuit output via sequestration of the core RNAP by the sigma factor.    

The model proposed here is a foundational one that may be easily expanded to include any number of genes.  Of course, any extension of the model would lead to increases in the dimensionality of the state and parameter spaces and bring it further into the regime of the well-known `parameter problem'  \cite{Gunawardena:2010fp}.   However,  the robustness of biological systems would lead us to suspect that any realistic biological model would not be particularly sensitive to specific values of parameters such as rate constants---indeed, order-of-magnitude approximations are often sufficient to explain and predict system behavior.   Should better parameter estimates be required, there exists a large and growing number of computational tools specifically designed for this purpose \cite{Ashyraliyev:2008p749}.  In addition,   technological advances in microfluidics and experimental platform miniturization are making high-throughput and quantitative measurements increasingly feasible; for example, the parallel characterization of a large number of independent biomolecular association and dissociation rates has recently been demonstrated \cite{Geertz:2012kr}.  Such computational and experimental methods are compatible with our modeling framework and TX-TL breadboard system.

 \addtolength{\textheight}{-1cm}   


\section*{APPENDIX}

\subsection{Methods} \label{methods}

Preparation of the TX-TL extract was as described previously~\cite{Shin:2010p1243, Shin:2012vj}.  The deGFP reporter construct was also described in that work. The noninteracting protein (TetR) was expressed from a $\mathrm{P_{LlacO-1}}$ regulatory part composed of a promoter specific to $\sigma^{70}$ flanked by two {\it lac} operators.  The secondary sigma factor $\sigma^{28}$ was expressed from an OR2-OR1-Pr regulatory part composed of a $\sigma^{70}$-specific promoter flanked by two lambda Cl operators.  The untranslated RNA gene was expressed off plasmid pAPA1256 from \cite{Lucks:2011hq}.  

Data were collected over two separate experimental runs using a Victor X3 plate reader set at 29$^{\circ}$C.  Measured fluorescence values were converted to concentrations using a predetermined calibration curve and plotted with an 8 minute offset to account for the time between the mixing of breadboard components and the start of data collection.  Simulations were done using Mathematica.

\begin{table}[!h]
\renewcommand{\arraystretch}{1.4}
\caption{Model parameters}
\vspace{-.4cm}
\label{parameterTable}
\centering
\hspace{-.9cm}
\begin{minipage}[t]{0.37\linewidth}
\vspace{0pt}
\centering
{\scriptsize
\begin{tabular}{c|l}
\hline
\bfseries Param. & \bfseries Value \\
\hline\hline
$k_{x,TX}$ & 0.05 s$^{-1}$ \\
\hline
$k_{xm,deg}$ & 0.0018 s$^{-1}$ \\
\hline
$k_{x,TL}$ & 0.05 s$^{-1}$  \\
\hline
$k_{\mathrm{X}+}$ & $3\!\times\!10^7$ M$^{-1}$s$^{-1}$  \\
\hline
$k_{\mathrm{X}-}$ &  6 s$^{-1}$  \\
\hline
$k_{x+}$ &  $3\!\times\!10^7$ M$^{-1}$s$^{-1}$ \\
\hline
$k_{x-}$ & 0.24 s$^{-1}$ \\
\hline
$L_{x, protein}$ & 800 bp \\
\hline
$L_{x,s2}$ & 800 bp \\
\hline
$V_{TX}$ &  3 bp$\cdot$s$^{-1}$  \\
\hline
$V_{TL}$ &  4 bp$\cdot$s$^{-1}$  \\
\hline
$k_{\mathrm{ES1}+}$ & $3\!\times\!10^7$ M$^{-1}$s$^{-1}$  \\
\hline
$k_{\mathrm{ES1}-}$ & $7.8\!\times\!10^{-3}$ s$^{-1}$\\
\hline
$[\mathrm{S1}]_\text{tot}$ & 30 nM  \\
\hline
$k_{mat}$ & 0.002 s$^{-1}$ \\
\hline
\end{tabular}  
}
\end{minipage}
\hspace{1cm}
\begin{minipage}[t]{0.37\linewidth}
\vspace{0pt}
\centering
{\scriptsize
\begin{tabular}{c|l}
\hline
\bfseries Param. & \bfseries Value \\
\hline\hline
$k_{y,TX}$ & 0.05 s$^{-1}$  \\
\hline
$k_{ym,deg}$ & 0.0018 s$^{-1}$  \\
\hline
$k_{y,TL}$ & 0.05 s$^{-1}$  \\
\hline
$k_{\mathrm{Y}+}$ & $3\!\times\!10^7$ M$^{-1}$s$^{-1}$  \\
\hline
$k_{\mathrm{Y}-}$ & 18 s$^{-1}$  \\
\hline
$k_{y+}$ &  $3\!\times\!10^7$ M$^{-1}$s$^{-1}$ \\
\hline
$k_{y-}$ & 0.48 s$^{-1}$ \\
\hline
$L_{x,RNA}$ & 90 bp \\
\hline
$L_{y}$ & 800 bp \\
\hline
$k_{\mathrm{ES2}+}$ & $3\!\times\!10^7$ M$^{-1}$s$^{-1}$ \\
\hline
$k_{\mathrm{ES2}-}$ &  $2.2\!\times\!10^{-2}$ s$^{-1}$ \\
\hline
$k_{s2,TL}$ & 0.05 s$^{-1}$  \\
\hline
$[\mathrm{R}]_\text{tot}$ & 1500 nM  \\
\hline
$[\mathrm{E}]_\text{tot}$ &  100 nM  \\
\hline
$[y]_\text{tot}$ & 2 nM  \\
\hline
\end{tabular}
}
\end{minipage}
\\
\end{table}

\subsection{Model parameters} \label{params}
The model parameter values used are listed in Table~\ref{parameterTable}.  Values were taken from \cite{Shin:2012vj,Karzbrun:2011bw}, and references therein, with the following exceptions and notes:
\begin{itemize}
\item Transcription rates $k_{i,TX}$ were assumed to be equal to (the previously-measured) $k_{y,TX}$.
\item Translation rates $k_{i,TL}$ were assumed to be equal to (the previously-measured) $k_{y,TL}$.
\item RNA degradation rates $k_{im,deg}$ were assumed to be equal to (the previously-measured) $k_{ym,deg}$.
\item When only a dissociation constant $K_d$ ($=k_-/k_+$) could be found, the on-rate ($k_+$) was taken to be $3\times10^7$ M$^{-1}$s$^{-1}$ (consistent with previous measurements of promoter association rates; see, e.g., \cite{Brunner:1987tt}), with the off-rate ($k_-$) set to $K_d \times k_+$.
\item $k_{y-}$ was fit using the `reporter only' data.
\item $k_{\mathrm{X}-}$ was fit using the `noninteracting protein' data.
\end{itemize}
For values previously given only as a range, reasonable values were taken from within the range. We note that in cell-free systems, the speeds of RNAP and ribosomes are slower than what has been measured {\it in vivo}.

\section*{ACKNOWLEDGMENT}

We thank Z. Sun, C. Hayes, J. Kim,  and E. Yeung for help with TX-TL experiments and useful discussions.  We also thank J. B. Lucks for the untranslated RNA gene construct.  This work is supported by the Institute for Collaborative Biotechnologies through grant W911NF-09-0001 from the U.S. Army Research Office, and the DARPA Living Foundries Program under Contract HR0011-12-C-0065.  The content of the information does not necessarily reflect the position or the policy of the Government, and no official endorsement should be inferred.
%
%


%
%
\bibliographystyle{IEEEtran}
\bibliography{ACC_resubmit}


\end{document}